\documentclass[conference]{IEEEtran}
\ifCLASSINFOpdf
  \usepackage[pdftex]{graphicx}
\else
\fi
\usepackage{amsmath}
\usepackage{xcolor}
\usepackage{color}
\usepackage{booktabs}
\usepackage{amssymb}

\usepackage{flushend}
\begin{document}
\addtolength{\topmargin}{0.02in}
\addtolength{\textheight}{-0.02in}
%
\title{Jamming Detection in 5G/6G Networks: From O-RAN Concept to OCUDU Deployment}

\author{
\IEEEauthorblockN{Marcin Hoffmann}
\IEEEauthorblockA{\textit{Rimedo Labs}}
\IEEEauthorblockA{\textit{Poznan University of Technology} \\
Poznan, Poland
}
\and
\IEEEauthorblockN{Lukasz Kulacz}
\IEEEauthorblockA{\textit{Rimedo Labs}}
\IEEEauthorblockA{\textit{Poznan University of Technology} \\
Poznan, Poland
}
\and
\IEEEauthorblockN{Osama Baldo}
\IEEEauthorblockA{\textit{Keysight Technologies} \\
Santa Rosa, California, United States
}
\and
\IEEEauthorblockN{Marcin Pakula}
\IEEEauthorblockA{\textit{Rimedo Labs}}
\IEEEauthorblockA{\textit{Poznan University of Technology} \\
Poznan, Poland
}
\and
\IEEEauthorblockN{Balaji Raghothaman}
\IEEEauthorblockA{\textit{Keysight Technologies} \\
Santa Rosa, California, United States
}
}


%


\maketitle

\begin{abstract}
RF jamming poses a severe threat to 5G/6G networks, increasing packet latency being especially disruption to mission-critical URLLC services. This paper introduces a proactive Jamming Detection xApp (JD-xApp) for the Open RAN architecture that detects attacks by monitoring the moving-average Block Error Rate (BLER) via the E2 interface. Upon detection, the algorithm overrides standard link adaptation, enforcing a robust upper ceiling on the Modulation and Coding Scheme (MCS) to stabilize latency. To counter O-RAN platform adoption challenges, we transition the framework to an open-source Centralized Unit/Distributed Unit implementation (OCUDU). Evaluated with high-end Keysight lab equipment including the UXM 5G Wireless Test Platform and the PROPSIM F64 channel emulator, the JD-xApp reduces the expected number of packet retransmission attempts by about 67.3\%. Finally, the system's feasibility is validated via over-the-air deployment using the POWDER lab infrastructure.
\end{abstract}


%
\IEEEpeerreviewmaketitle

\section{Introduction}

The 5G/6G networks offer high security and trustworthiness at the level of protocols and authentication. However, due to the open nature of the wireless channel, they are still prone to relatively straightforward attacks on the Radio Access Network (RAN), such as Radio Frequency (RF) jamming~\cite{Pirayesh2022, kryszkiewicz2023}. In essence, access to the radio channel is restricted by formal regulations, agreements, and protocols, yet this does not prevent their intentional abuse. Moreover, jamming harmful to 5G/6G networks can be realized using cheap off-the-shelf components and does not require sophisticated algorithms to significantly reduce users' Quality of Service (QoS). For example, a keyed jammer changing its state (on/off) faster than Channel State Information (CSI) reports can disrupt scheduler Modulation and Coding Schemes (MCS) allocation, increasing packet latency~\cite{bogucka2025}. This can have significant negative consequences for both military and civilian mission-critical Ultra-Reliable Low-Latency Communication (URLLC) services. The key issue with a jamming attack is that it can only be truly overcome by eliminating the interference source, which requires its localization and action taken by the proper public services. Although this usually requires a significant amount of time, some temporal actions can be taken to reduce its negative impact on the 5G/6G network, e.g., the carrier frequency can be adjusted so as to avoid the jammed band~\cite{jia20222}. While this idea seems reasonable for a single cell, in practice it might be very challenging to coordinate relocation of carrier frequencies within a broader area of the mobile network, and potentially between multiple operators. As an alternative approach, which does not require reorganization of spectrum allocation in 5G/6G networks, the effects of jamming, such as latency, can be reduced by taking control over the scheduler and setting fixed MCS~\cite{bogucka2025}. This enables overcoming scheduler instability when the keyed jamming rate exceeds CSI reporting intervals. 

However, typically RAN is provided by a single vendor and does not expose both low-layer Key Performance Metrics (KPMs) and the capability of interaction with the scheduler. This issue is addressed by the concept of Open RAN, which standardizes open interfaces that, together with the RAN Intelligent Controller (RIC), allow the deployment of third-party algorithms in the form of xApps (for near-real-time optimization) and rApps (for non-real-time optimization)~\cite{hoffmann2024}. In particular, the E2 interface between the Near Real-Time RIC (Near-RT RIC), which comes with a Lower Layer Control Service Model (E2SM-LLC), supports interaction with the scheduler~\cite{oran_e2sm_llc}. Unfortunately, the practical adoption of Near-RT RIC and xApps is limited to mostly open-source platforms, implementing only a small portion of the O-RAN ALLIANCE specifications~\cite {rodgers2025xapp}. Moreover, despite referring to the O-RAN ALLIANCE specifications, they usually come with different API and SDKs, which introduces additional software development overhead and limits true interoperability~\cite{hoffmann2024}. From this perspective, the Open RAN concept can evolve in a way that, to deploy a third-party algorithm, the gNB should expose any kind of proprietary API enabling the exposure of KPMs' and control capabilities. Recently, the leading candidate for such a deployment is the OCUDU, which is part of the Linux Foundation and is an open-source implementation of the O-RAN Centralized Unit and Distributed Unit originating from srsRAN~\cite{ocudu_software}.

In this paper, in Sec.~\ref{sec:core_approach}, we introduce the Jamming Detection xApp (JD-xApp), which detects the attack through monitoring of Block Error Rate (BLER) and mitigates its negative effects on latency through modification of the MCS allocated to the user. The provided insights go beyond the brief description in~\cite{bogucka2025}. Then, in Sec.~\ref{sec:ocudu_migration}, we discuss how the initial O-RAN JD-xApp can be deployed on the OCUDU. In Sec.~\ref{sec:keysight_test}, the proposed JD-xApp is evaluated using high-end lab equipment provided by Keysight, including the UXM 5G Wireless Test Platform and the PROPSIM F64 channel emulator. Finally, recent tests of JD-xApp deployment on OCUDU are presented in Sec.~\ref{sec:ocudu_migration}. These are done over-the-air using the infrastructure of a POWDER lab~\cite{BREEN2021108281}.

\section{Jamming Detection xApp - Core Approach} \label{sec:core_approach}
\label{jd-core-approach}

The proposed JD-xApp provides a proactive, closed-loop framework for maintaining low-latency communication in 5G networks under intentional RF interference. Operating as an xApp on the O-RAN Near-RT RIC, the algorithm continuously monitors downlink transport block reception statistics (ACK/NACK signals) collected over the E2 interface using E2SM-LLC~\cite{oran_e2sm_llc}. By calculating the moving-average BLER over a sliding observation window, the JD-xApp is able to detect both continuous and pulsed jamming attacks.

\begin{figure}[!h]
\centering
\includegraphics[width=0.47\textwidth]{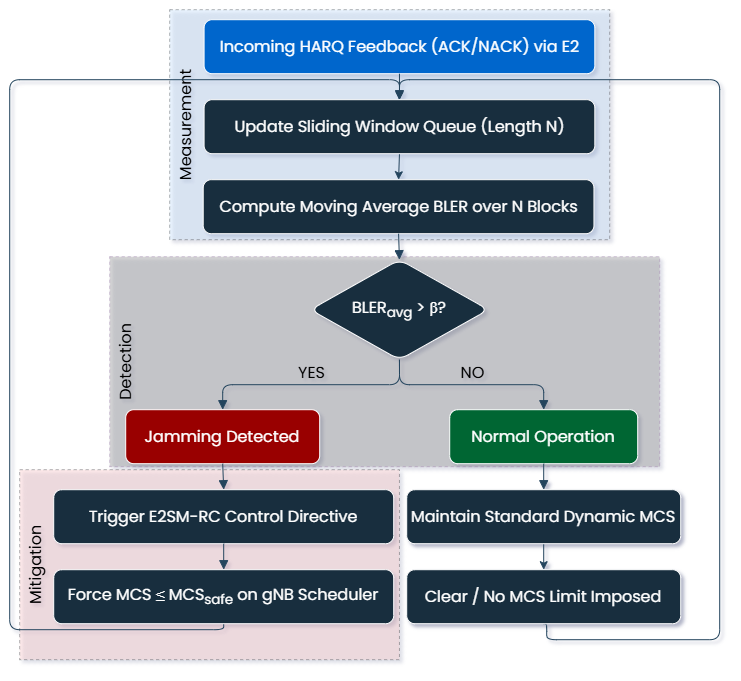}
\caption{Jamming detection and mitigation: core approach.}
\label{jd-alg}
\end{figure}

Once a jamming event is detected, the mitigation mechanism instantly overrides standard link adaptation by issuing real-time control policies to the gNB MAC scheduler. Using E2SM-LLC, it enforces a strict upper ceiling on the Modulation and Coding Scheme (MCS), forcing transmission down to an ultra-robust profile. Although this administrative capping reduces overall spectral efficiency and throughput, it drastically lowers transport block decoding errors under elevated noise floors. By reducing heavy retransmission delays and buffer stalls, the algorithm stabilizes round-trip latency and prevents link failures for delay-sensitive traffic. The core end-to-end detection and mitigation workflow of JD-xApp is shown in Fig.~\ref{jd-alg}.

\subsection{Measurement Telemetry Aggregation}
The UE transmits Hybrid Automatic Repeat Request (HARQ) feedback (ACK/NACK) over the Physical Uplink Control Channel (PUCCH) or Physical Uplink Shared Channel (PUSCH) in response to downlink Transport Blocks (TBs). The E2 Node (O-DU) aggregates these transport block reception outcomes and, via the E2 Application Protocol (E2AP) using the E2SM-LLC, streams periodic or event-triggered indication messages containing per-UE ACK/NACK statistics to the Near-RT RIC.

\subsection{Detection Algorithm Execution (Near-RT RIC)}

The JD-xApp processes incoming transport block statuses (ACK/NACK), marked as $\mathrm{TB}_i$, within a sliding observation window of $N$ consecutive transmissions. The moving average BLER for a given UE is computed as:

\begin{equation}
\mathrm{BLER}_{\mathrm{avg}} = \frac{1}{N} \sum_{i=1}^{N} \mathbb{I}(\mathrm{TB}_i = \mathrm{NACK}),
\label{eq:bler_avg}
\end{equation}
where $\beta$ is a calibrated detection threshold.
If $\mathrm{BLER}_{\mathrm{avg}} > \beta$, the JD-xApp flags a jamming attack.

\subsection{Mitigation Directive Enforcement}

Upon jamming detection, standard gNB link adaptation is suspended for a specific period of time, since rapid channel degradation from jammer bursts typically causes conventional adaptive modulation to fail due to delayed Channel State Information (CSI) reports. The JD-xApp issues an E2 Control message via the E2SM-LLC, imposing an absolute upper ceiling on the modulation scheme ($\mathrm{MCS} \le \mathrm{MCS}_{\mathrm{safe}}$, e.g., $\mathrm{MCS} = 2$, corresponding to robust BPSK/QPSK with heavy coding) directly onto the E2 Node's MAC scheduler. This forced ultra-robust modulation allows transport blocks to withstand elevated noise floors and interference bursts, eliminating excessive HARQ retransmission delays for latency-critical URLLC traffic. 

\section{O-RAN Concept to OCUDU Deployment} \label{sec:ocudu_migration}

While O-RAN E2SM-LLC is a good fit for deployment of JD-xApp in the sense of providing a unified mechanism to obtain the input KPMs and MCS control capability, the key issue is its poor adoption by existing RIC platforms; e.g., see available E2SMs in~\cite{rodgers2025xapp,hoffmann2024}. Moreover, the E2SM-LLC utilizes ASN1-encoded messages, which introduces additional development overhead. In contrast, the OCUDU provides an open-source implementation of a 5G gNB base station~\cite{ocudu_software}. It does not require complicated ASN1 syntax and exposes metrics and control capabilities directly through WebSocket using a standard JSON format. The data can be reported at 500~ms intervals, including various operational levels, and goes beyond O-RAN specifications. For example, the Radio Link Control layer provides per-UE Data Radio Bearer metrics, such as maximum PUD latency. At the MAC layer, the telemetry captures scheduler performance, such as allocation error counters or real-time bitrates. At the Distributed Unit layer, the OCUDU evaluates allocated MCS, BLER, and Signal-to-interference-plus-Noise Ratio (SINR). The metrics also consist of the Centralized Unit Control Plane, with active PDU sessions, and the status of N2/NGAP interface. Finally, the Centralized Unit User Plane consists of PDCP layer metrics for Dowlink or Uplink traffic, and tracking average throughput. Moreover, using the OCUDU API, it is possible to enforce MCS for a certain UE. 

From this perspective, OCUDU provides all necessary metrics to deploy JD-xApp. The key observation is that the input and output data, along with the core approach, are literally the same. The major change is in the interface for communication with RAN. The comparison between O-RAN E2SM-LLC-based  and OCUDU approaches is presented in Tab.~\ref{tab:ocudu_vs_oran}.   
\begin{table}[htbp]
\centering
\caption{Comparison between O-RAN and OCUDU \\ deployment of JD-xApp.}
\label{tab:ocudu_vs_oran}
\begin{tabular}{lrr}
\toprule
\textbf{} & \textbf{O-RAN (Near-RT RIC)} & \textbf{OCUDU}  \\
\midrule
Interface             & E2SM-LLC & Proprietary API  \\
Format             & ASN1 & JSON   \\
KPM: ACK/NACK              & Yes & Yes   \\
Control: MCS             & Yes  & Yes   \\
Market Adoption      & Poor  & Growing    \\
\bottomrule
\end{tabular}
\end{table}

\section{Tests of JD-xApp Core Algorithm} \label{sec:keysight_test}

As the JD-xApp core approach is agnostic of the platform used for its deployment, the key requirement is to validate it using reliable tools. To achieve that, we used high-end lab equipment provided by Keysight Technologies:
\begin{itemize}
    \item \textbf{E7515B UXM 5G Wireless Test Platform}, enabling emulation of the 5G New Radio protocol stack. We use it to set up the communication between the cell and UE being subject to a jamming attack.
    \item \textbf{F8800A PROPSIM F64 Radio Channel Emulator}, enabling realistic and repeatable radio link emulation with dynamic introduction of Additive White Gaussian Noise (AWGN). We used it for modeling radio channel characteristics between the UE and the cell, with AWGN being used to mimic a jamming attack.
\end{itemize}
\subsection {BLER-Based Link Adaptation Test}
\begin{figure}[!h]
\centering
\includegraphics[width=0.49\textwidth]{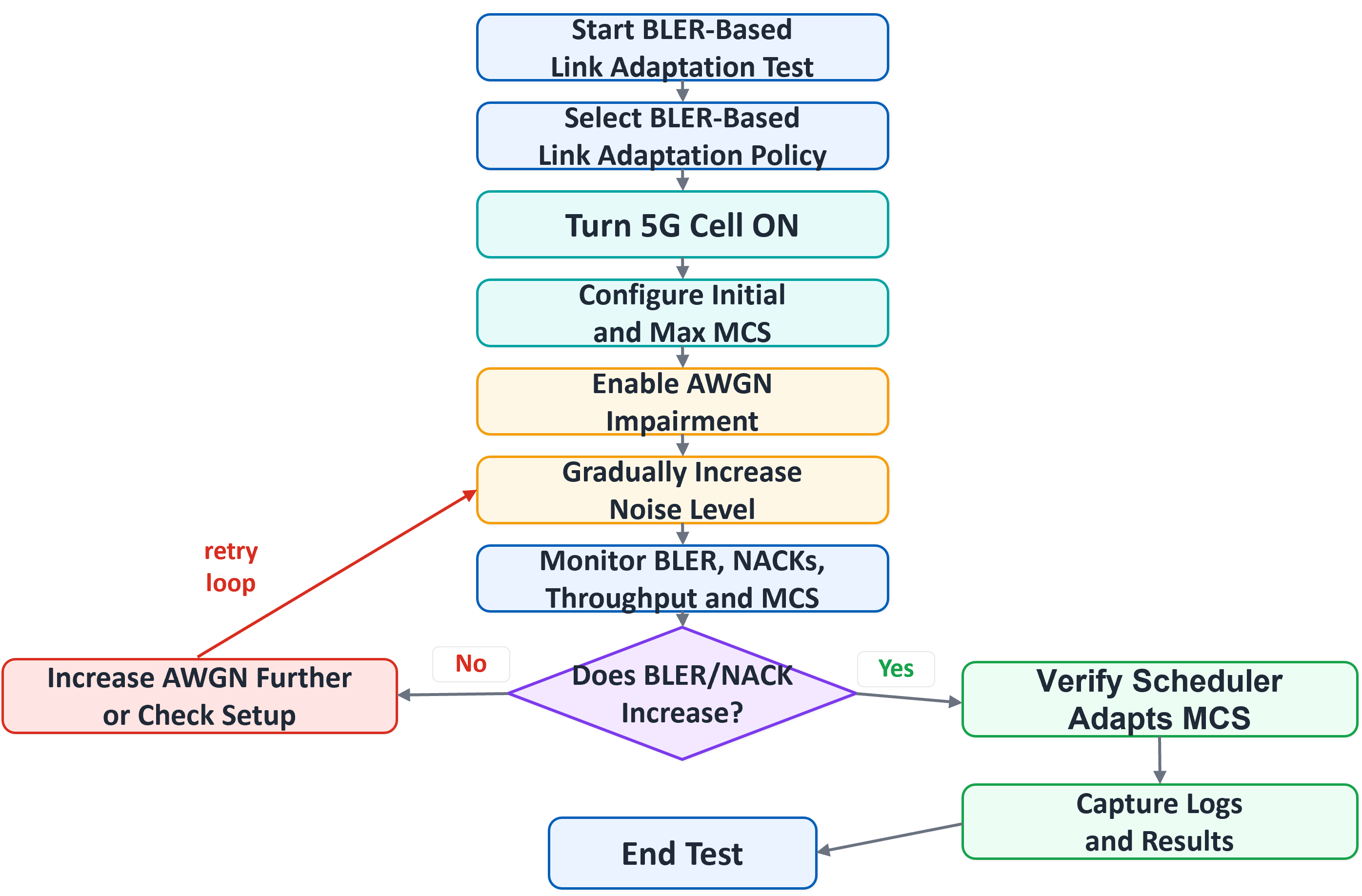}
\caption{Log processing and metric extraction.}
\label{fig:bler_based}
\end{figure}

To generate data for evaluation of JD-xApp using the Keysight UXM 5G Wireless Test Platform, we followed the BLER-Based Link Adaptation Test, which is summarized in Fig.~\ref{fig:bler_based}. After starting the test a link-adaptation policy is selected. It can be either default MCS selection based on CSI, which reflects the standard operation of the gNB, or a fixed MCS, corresponding to the JD-xApp mitigation mechanism. Then, the 5G cell emulated by UXM is started and configured with the initial MCS according to the policy. During the test, a jamming attack is modeled by the PROPSIM F64 channel emulator by introducing AWGN to the radio link, which causes packet retransmissions indicated by a growing BLER. To achieve sufficient deterioration of propagation conditions caused by jamming, the AWGN is increased iteratively and, after reaching a certain BLER level, MCS adaptation in the scheduler is verified, and logs are dumped to files for further processing. 

\subsection{Log Processing and Metric Extraction}

To validate the detection and mitigation logic under realistic radio conditions, the system was evaluated using log traces generated from a Keysight UXM 5G Wireless Test Platform. These logs are extensive, capturing a wide range of link-layer behaviors and statistics beyond the parameters required for jamming detection, and are streamed as sequential text records rather than structured data.

\begin{figure}[!h]
\centering
\includegraphics[width=0.40\textwidth]{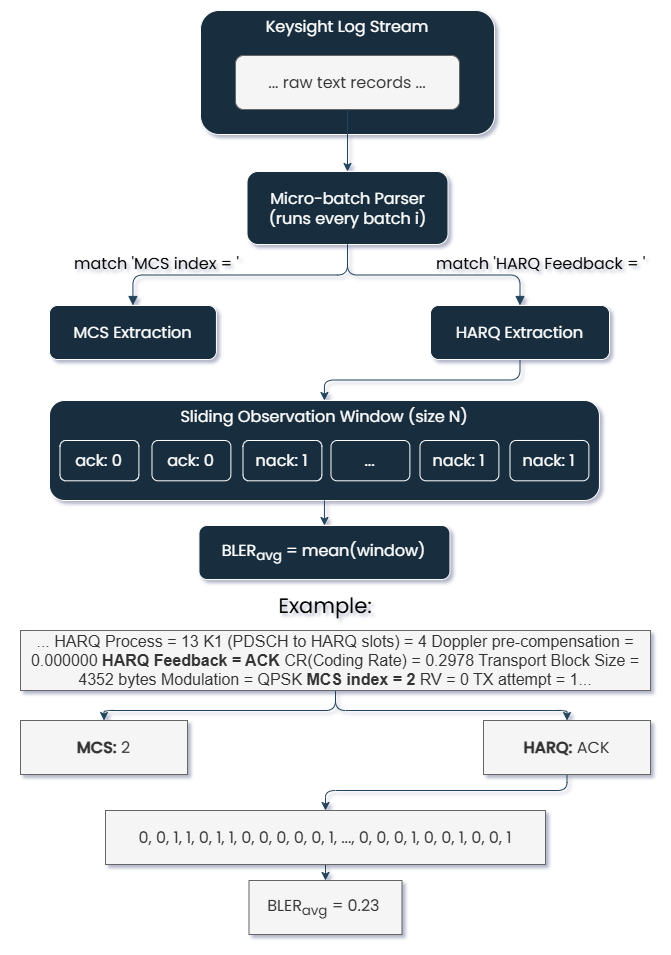}
\caption{Log processing and metric extraction.}
\label{keysight-data}
\end{figure}

A dedicated parsing routine, shown of Fig.~\ref{keysight-data}, processes these records in batches extracting two categories of metrics relevant to the proposed algorithm:

\begin{itemize}
    \item \textbf{HARQ Feedback:} Each log entry containing a \texttt{HARQ Feedback} field is scanned to determine whether the reported outcome is a positive acknowledgment (ACK) or a negative acknowledgment (NACK). The routine maintains running counters for total feedback events and successful acknowledgments, while appending a binary indicator (0 for ACK, 1 for NACK) to a time-ordered sequence. This sequence directly corresponds to the indicator function $\mathbb{I}(\mathrm{TB}_i = \mathrm{NACK})$ used in the $\mathrm{BLER}_{\mathrm{avg}}$ equation~(\ref{eq:bler_avg}).
    \item \textbf{MCS Index:} Each log entry containing an \texttt{MCS index} field is parsed to extract the corresponding integer value. The routine accumulates a running total and occurrence count to support average MCS computation, and additionally appends each observed value to a time-ordered sequence for downstream analysis of link adaptation behavior over the course of the experiment.
\end{itemize}

By aggregating these metrics incrementally across batches rather than loading the full log into memory at once, the parser accommodates the scale of the Keysight-generated traces while preserving the temporal ordering necessary to reconstruct the sliding-window BLER calculation offline, mirroring the online behavior of the JD-xApp described in Sec.~\ref{jd-core-approach}.

\subsection{Experimental Results}

The proposed algorithm was evaluated against a pulsed jamming scenario generated with a Keysight UXM 5G Wireless Test Platform attached to the PROPSIM F64, in which the interference source was active from approximately $t=175\,\mathrm{s}$ to $t=405\,\mathrm{s}$. To isolate the contribution of the JD-xApp, two configurations were logged from the same experimental scenario: a \emph{reference} run with the mitigation algorithm disabled (standard gNB link adaptation only), and an \emph{active} run with the JD-xApp enabled.

\begin{figure}[htbp]
    \centering
    \includegraphics[width=0.9\linewidth]{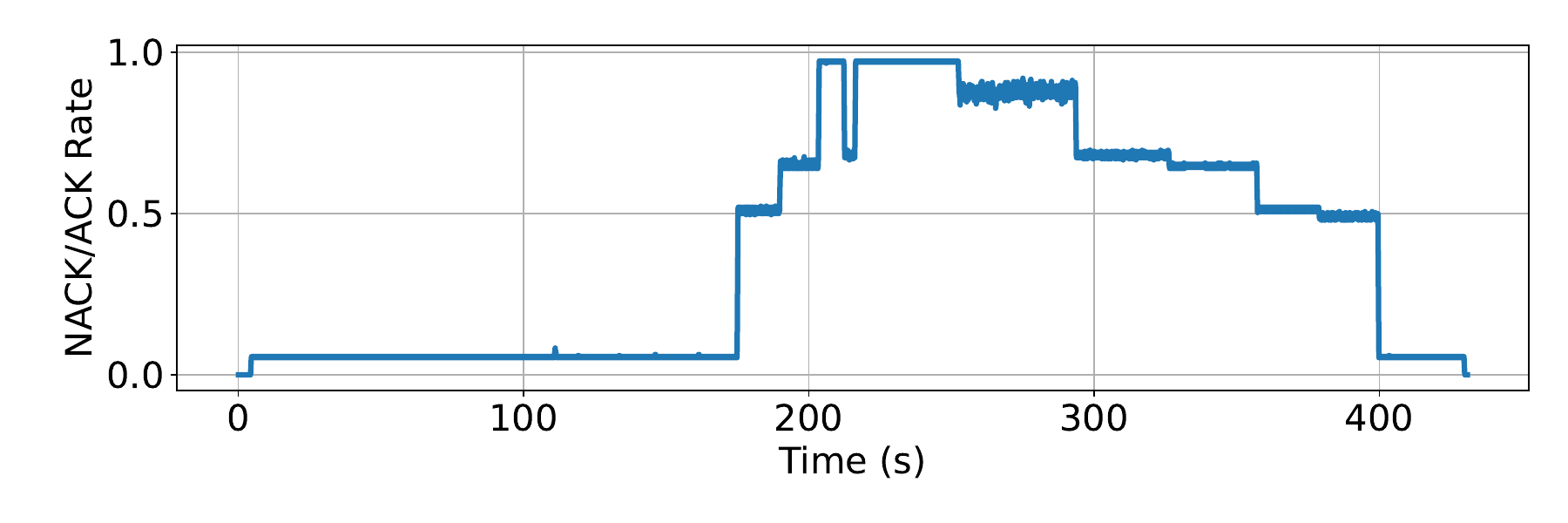}
    \caption{NACK/ACK rate over time with the mitigation algorithm \emph{disabled} (reference run). The rate remains persistently elevated throughout the jamming interval ($\approx175$--$405\,\mathrm{s}$).}
    \label{fig:nack_ref}
\end{figure}

\begin{figure}[htbp]
    \centering
    \includegraphics[width=0.9\linewidth]{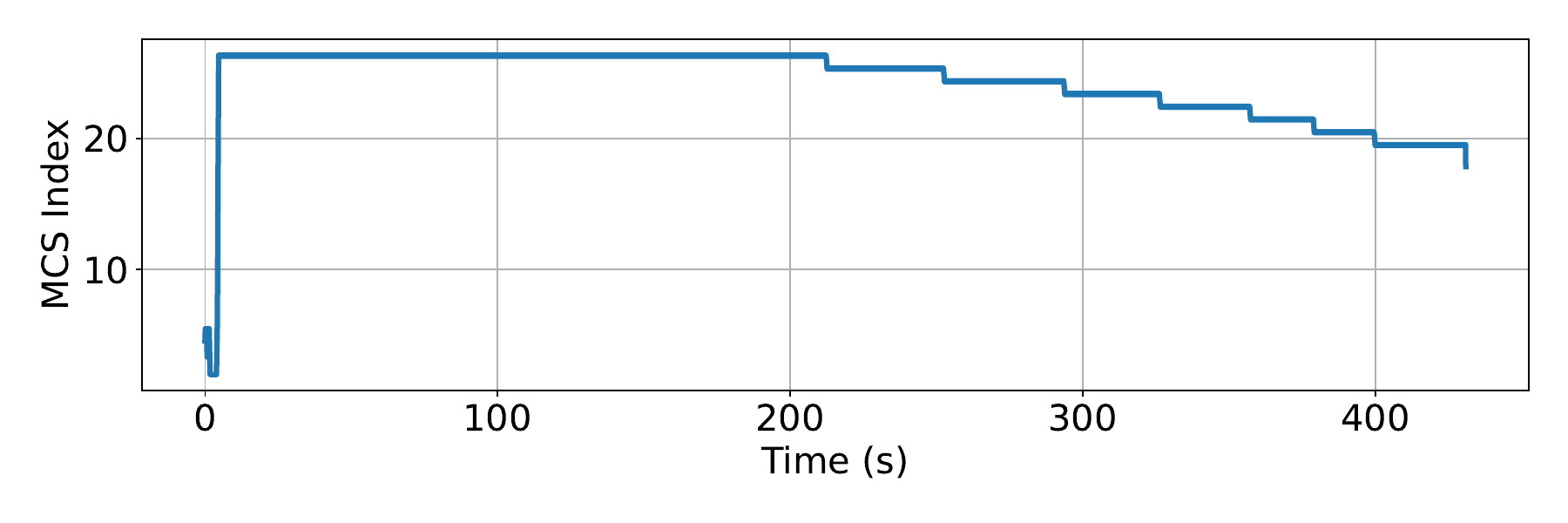}
    \caption{MCS index over time with the mitigation algorithm \emph{disabled} (reference run). Standard link adaptation fails to react promptly to the onset of jamming.}
    \label{fig:mcs_ref}
\end{figure}

Fig.~\ref{fig:nack_ref} and Fig.~\ref{fig:mcs_ref} show the NACK/ACK rate and MCS index for the reference run. During the jamming interval, the NACK/ACK rate rises sharply and remains persistently elevated, reaching values close to $1.0$ for extended periods. Critically, the standard MCS controller reacts only sluggishly to this degradation: the MCS index remains pinned near its maximum value ($\mathrm{MCS}\approx27$) for most of the jamming duration and only begins a gradual decline towards the end of the interference window Fig.~\ref{fig:mcs_ref}. 

\begin{figure}[htbp]
    \centering
    \includegraphics[width=0.9\linewidth]{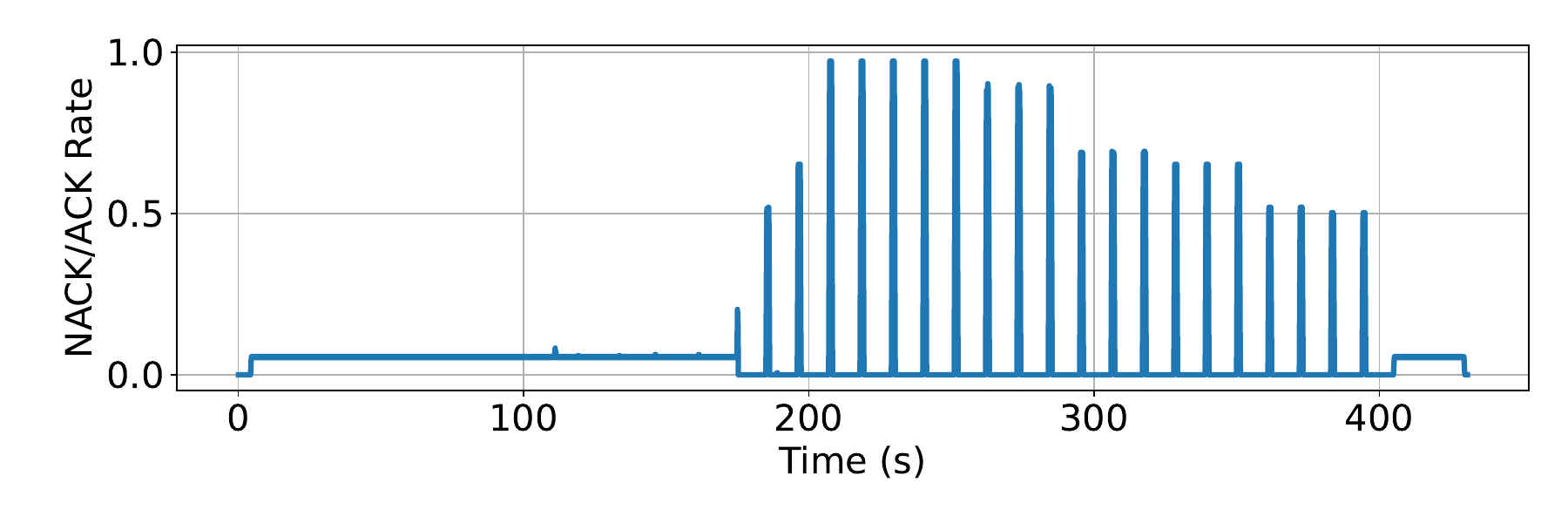}
    \caption{NACK/ACK rate over time with the JD-xApp \emph{enabled}. Error bursts are quickly suppressed, appearing as short spikes rather than sustained degradation.}
    \label{fig:nack_alg}
\end{figure}

\begin{figure}[htbp]
    \centering
    \includegraphics[width=0.9\linewidth]{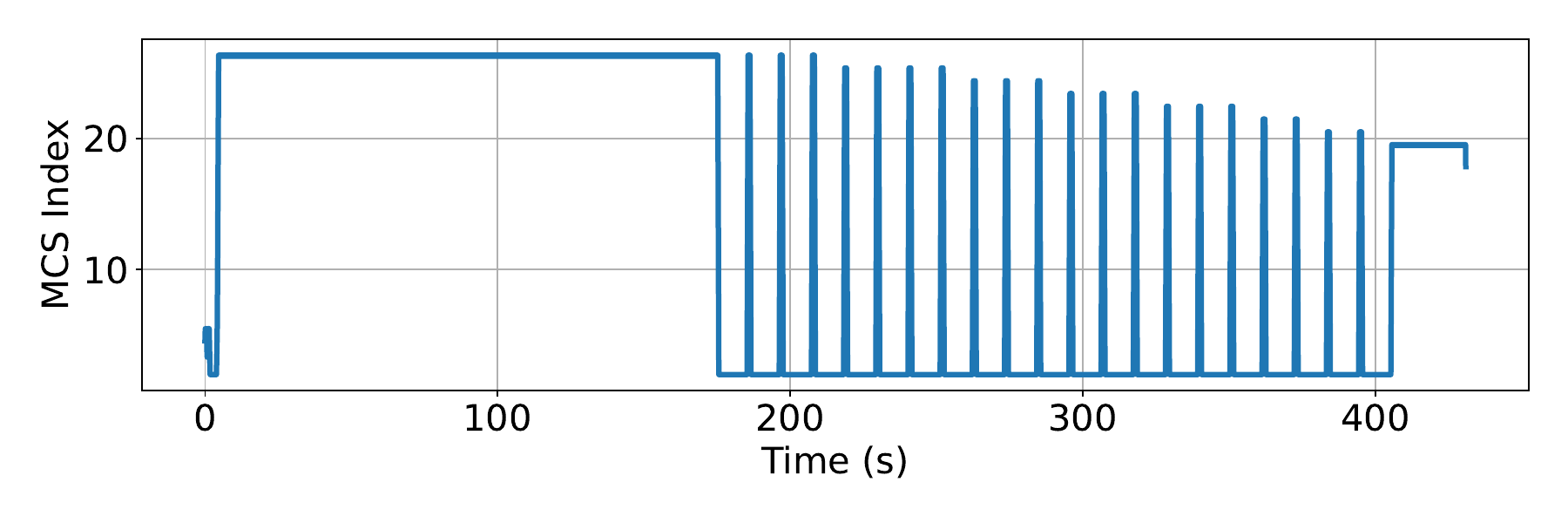}
    \caption{MCS index over time with the JD-xApp \emph{enabled}. The controller repeatedly forces the MCS down to $\mathrm{MCS}_{\mathrm{safe}}$ upon jamming detection and releases it once conditions improve.}
    \label{fig:mcs_alg}
\end{figure}

Fig.~\ref{fig:nack_alg} and Fig.~\ref{fig:mcs_alg} show the same metrics with the JD-xApp active. The NACK/ACK rate exhibits a markedly different pattern: rather than remaining persistently high, it appears as a series of short, sharp spikes, each promptly pulled back down towards zero. This behavior is explained by the corresponding MCS trace, which shows the xApp repeatedly and rapidly forcing the MCS index down to its safe floor ($\mathrm{MCS}_{\mathrm{safe}}\approx2$) at the onset of each detected error burst, before allowing it to recover towards nominal values once the assumed time.

\begin{figure}[htbp]
    \centering
    \includegraphics[width=0.9\linewidth]{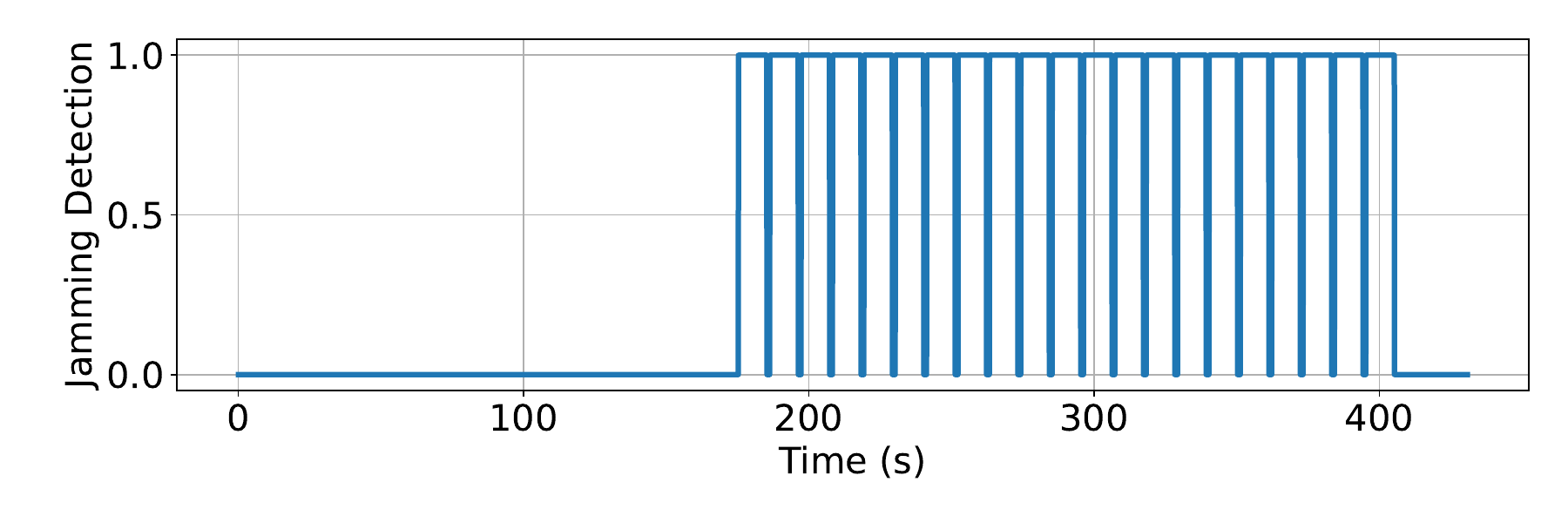}
    \caption{Binary jamming-detection flag output by the JD-xApp, correlating with the MCS-capping events.}
    \label{fig:jamming}
\end{figure}

Fig.~\ref{fig:jamming} shows the corresponding binary jamming-detection flag produced by the JD-xApp, which toggles to $1$ in near-perfect correspondence with the MCS-capping events in Fig.~\ref{fig:mcs_alg}, confirming that the mitigation actions are correctly triggered by the detection logic.

Overall, these results indicate that while the reference (uncontrolled) system suffers from a prolonged period of severe link degradation under jamming, the proposed closed-loop mechanism trades a temporary, self-limiting reduction in MCS for a substantial reduction in sustained block error rate, consistent with the design goal of prioritizing link robustness over peak throughput during interference events.

\subsection{Impact on Retransmission Overhead and Latency}

To quantify the latency benefit of the proposed mitigation mechanism, the number of transmission attempts required per downlink transport block was extracted from the same paired experiment (mitigation disabled vs. enabled), over $N = 415{,}972$ logged transport blocks in each run. Under HARQ operation, a transport block that is acknowledged on the first attempt incurs a nominal transmission time $T$, while each additional retransmission attempt incurs a further delay $T_{\mathrm{extra}}$ due to HARQ round-trip and scheduling overhead. The expected per-block latency is therefore

\begin{equation}
\mathbb{E}[\mathrm{Latency}] = T + \big(\mathbb{E}[\mathrm{attempts}] - 1\big) \, T_{\mathrm{extra}}
\label{eq:expected_latency}
\end{equation}

Table~\ref{tab:retransmissions} summarizes the observed attempt-count distribution for both configurations and the resulting expected latency overhead, expressed in multiples of $T_{\mathrm{extra}}$.

\begin{table}[htbp]
\centering
\caption{Transmission attempt distribution and resulting latency overhead, with and without the jamming mitigation algorithm.}
\label{tab:retransmissions}
\begin{tabular}{lrr}
\toprule
\textbf{Metric} & \textbf{Reference (no alg.)} & \textbf{With JD-xApp}  \\
\midrule
$P(1^{\mathrm{st}}\text{ attempt})$              & 62.16\% & 94.01\%  \\
$P(2^{\mathrm{nd}}\text{ attempt})$             & 20.38\% & 4.23\%   \\
$P(3^{\mathrm{rd}} \text{ attempt})$             & 11.79\% & 1.15\%   \\
$P(4^{\mathrm{th}} \text{ attempt})$             & 5.67\%  & 0.61\%   \\
$\mathbb{E}[\text{attempts}]$      & 1.610   & 1.084    \\
$\mathbb{E}[\text{extra delay}]$  & 0.610 & 0.084  \\
\bottomrule
\end{tabular}
\end{table}

Without mitigation, $37.84\%$ of transport blocks required at least one retransmission, and the expected number of attempts per block reached $1.61$, meaning that on average each block incurred $0.61\,T_{\mathrm{extra}}$ of additional delay beyond the nominal transmission time $T$. With the JD-xApp active, the probability of requiring any retransmission dropped to $5.99\%$, and the expected number of attempts fell to $1.084$, corresponding to only $0.084\,T_{\mathrm{extra}}$ of expected additional delay — an $86.3\%$ reduction in expected retransmission-induced latency per transport block.

Aggregated over the full experiment, this corresponds to approximately $253{,}600$ retransmission attempts in the reference run versus approximately $34{,}800$ retransmission attempts with mitigation enabled — a reduction of roughly $218{,}800$ retransmissions, or $86.3\%$ fewer HARQ retries overall. Since each avoided retransmission removes one full $T_{\mathrm{extra}}$ round-trip from the corresponding block's delivery time, this reduction directly translates into lower and more predictable round-trip latency, which is consistent with the goal of preserving delay-sensitive (URLLC) traffic performance under active jamming.

\section{Over the Air Tests on OCUDU} \label{sec:powder_test}

After the verification of the JD-xApp core approach using high-end Keysight UXM 5G Wireless Test Platform, we move toward its deployment on the OCUDU. Currently, we have integrated the JD-xApp mitigation part into OCUDU and use the infrastructure of the POWDER lab (see~\cite{BREEN2021108281}) to perform over-the-air tests. The topology for the evaluation scenario consists of the following elements:
\begin{itemize}
    \item \textbf{CN5G}: Hosts the Open5GS 5G Core Network.
    \item \textbf{OCUDU}: Serves as a combined Central Unit and Distributed Unit (CU/DU)
    \item \textbf{Sdru-SDR (USRP X310)}: Acts as a Software-Defined Radio (SDR) connected to the CU/DU. 
     \item \textbf{UE1, UE2}: Commercial User Equipments.
      \item \textbf{Uemon-SDR}: Provides monitoring and management capabilities across the radio-link setup.
      \item \textbf{JD-xApp}: Communicates with OCUDU using WebSocket and performs jamming detection based on the captured KPMs.
\end{itemize}
At the radio layer, OCUDU was configured to operate in 3GPP NR Band n78 with a channel bandwidth of 20 MHz and a subcarrier spacing of 30 kHz. The carrier frequency was defined as 3489.42 MHz. The USRP X310 SDR was set as a Radio Unit (RU). The RF gains were set at 30~dB for transmission and 20~dB for reception, operating at a baseband sampling rate of 92.16~MHz. 
A programmable attenuation matrix was used to control the Radio Frequency (RF) conditions between the network components and UEs. The prepared experimental environment therefore provided a platform for testing the developed algorithm in a 5G network. 
The jamming was introduced by rapidly switching the attenuation of the selected RF path between an attenuation level of $95~dB$ and the normal operating level. After the initial verification, the switching period was set to $10~ms$. This is synchronized with a 5G frame duration and CSI reports, meaning that after the OCUDU scheduler receives information about the radio channel, its state rapidly changes, causing BLERs due to MCS selection mismatch.

\begin{figure}[htbp]
    \centering
    \includegraphics[width=0.9\linewidth]{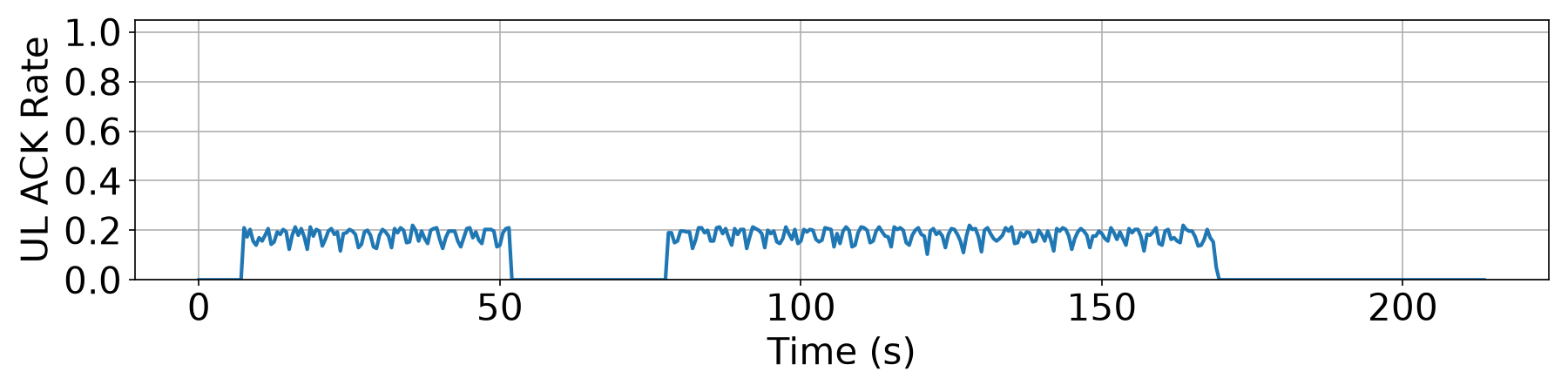}
    \caption{NACK/ACK rate over time. The rate remains persistently elevated throughout the jamming interval}
    \label{fig:powder-nack}
\end{figure}

This is clearly visible in Fig.~\ref{fig:powder-nack}, which shows the UL NACK/ACK ratio during the experiment. During the jamming attack the increases from close to 0\% to about 20\%, indicating that the default MCS allocation scheme fails to keep up with the rapidly degraded radio link. It corresponds to an increased number of unsuccessful transmissions, which requires retransmissions and introduces latency. Note that, for some sensitive URLLC services, even a small additional delay can be harmful.

\begin{figure}[htbp]
    \centering
    \includegraphics[width=0.9\linewidth]{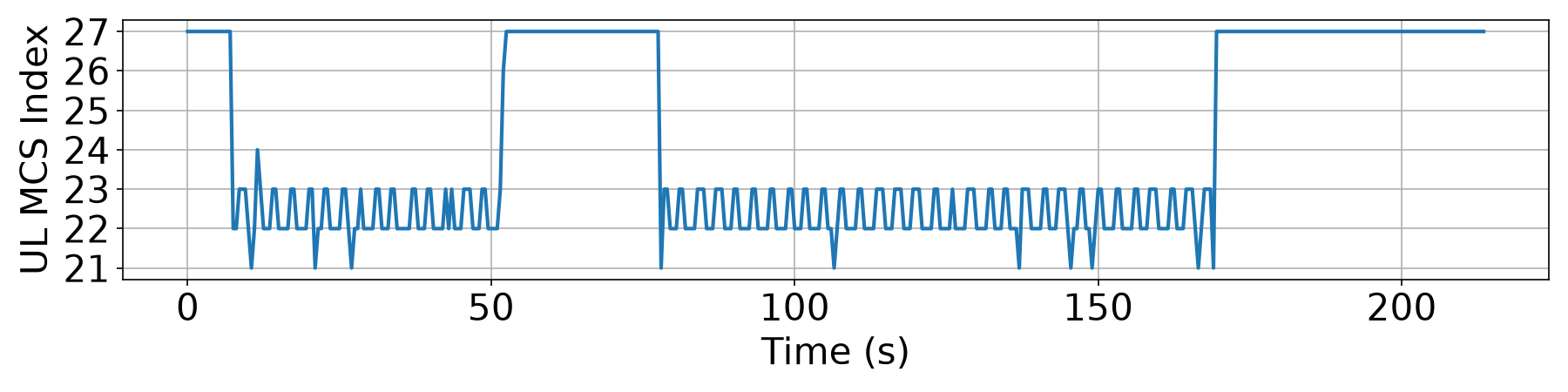}
    \caption{MCS index over time with jamming enabled.}
    \label{fig:powder-mcs}
\end{figure}

The related MCS index over time is presented in Fig.~\ref{fig:powder-mcs}. When jamming is enabled, the MCS decreases and fluctuates around $22-23$ values, reflecting the rapidly changing and degraded channel conditions. However, it is not enough to reduce BLER to a satisfactory level. This would require extending the OCUDU deployment of JD-xApp with a mitigation part, being responsible for adjusting the MCS during the attack.

\begin{figure}[htbp]
    \centering
    \includegraphics[width=0.9\linewidth]{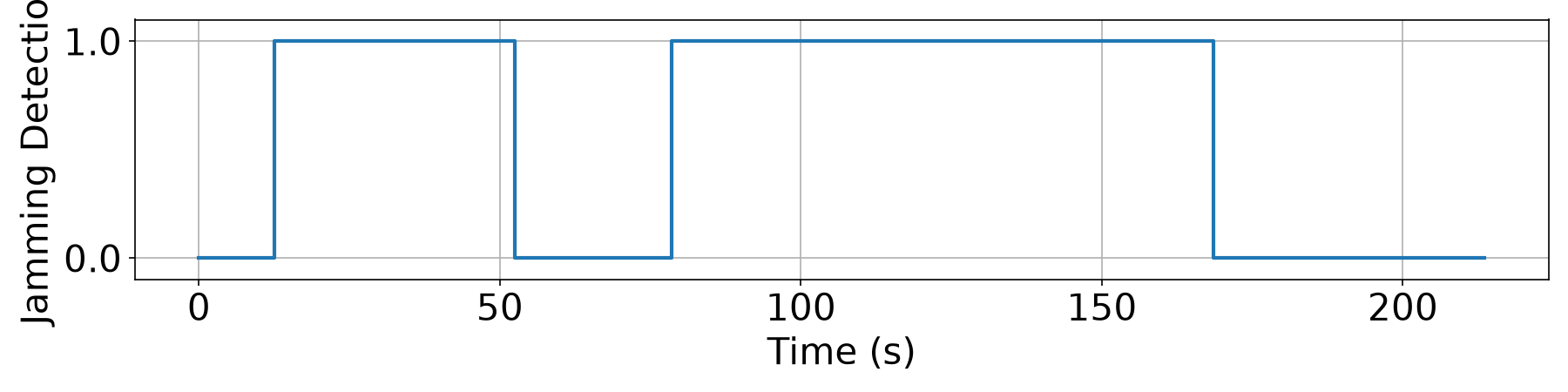}
    \caption{Jamming-detection flag output by the JD-xApp.}
    \label{fig:powder-xapp}
\end{figure}

Finally, Fig.~\ref{fig:powder-xapp} shows the jamming-detection flag generated by the JD-xApp. The detected intervals correspond closely to the periods of MCS degradation, indicating that the proposed detection mechanism is capable of reliably identifying the introduced jamming conditions.

\section{Conclusion} \label{sec:conclusions}
The keyed jamming of a rate higher than the CSI report intervals can result in an MCS allocation mismatch at the level of the MAC scheduler, resulting in high BLER and latency. This can be mitigated by the proposed JD-xApp, which enforces fixed MCS allocation during the jamming attack to reduce packet retransmissions. Evaluation of the JD-xApp core approach with the high-end Keysight UXM 5G Wireless Test Platform showed that, under jamming, it enables transmitting about 94\% of packets in the first attempt compared to only 62.16\% under default MCS allocation. Finally, the JD-xApp designed for O-RAN can be deployed on OCUDU, as we demonstrated through over-the-air tests in the POWDER lab. The OCUDU deployment will be subject to further development effort aiming to be extended with a mitigation part and extensive evaluation in terms of a broader set of KPMs.


\section*{Acknowledgment}

The authors would like to thank Keysight Technologies for support in collecting evaluation data with their lab equipment: a UXM 5G Wireless Test Platform and PROPSIM F64 channel emulator; and the POWDER lab managed by the University of Utah for sharing their infrastructure for OCUDU-based deployment. The work was supported by Polish Ministry of Science and Higher Education under grant number: 0312/SBAD/8169, and 0312/SBAD/8170.



%
\bibliography{bibliography} 
\bibliographystyle{IEEEtran}

\end{document}